\documentclass[twocolumn]{aastex631}
\usepackage{amsmath}
\usepackage{graphicx}
\usepackage{multirow}
\usepackage{hyperref}
\usepackage{float}
\usepackage{booktabs}
\usepackage{placeins}
\usepackage{longtable}
\usepackage{chemformula}
\usepackage[utf8]{inputenc}

\begin{document}

\title{Direct Evidence for Outflow Driven by Wolf--Rayet Stars in the Nearby Galaxy PGC~44685}
\author{Qian Huang}
\affiliation{School of Astronomy and Space Science, 
Nanjing University, Nanjing, Jiangsu 210093, China}
\affiliation{Key Laboratory of Modern Astronomy and Astrophysics (Nanjing University), Ministry of Education, Nanjing 210093, China}

\author{Zhengyi Chen}
\affiliation{National Astronomical Observatory of Japan, 2-21-1 Osawa, Mitaka, Tokyo 181-8588, Japan}
\affiliation{School of Astronomy and Space Science, 
Nanjing University, Nanjing, Jiangsu 210093, China}
\affiliation{Key Laboratory of Modern Astronomy and Astrophysics (Nanjing University), Ministry of Education, Nanjing 210093, China}

\author{Min Bao}
\affiliation{School of Physics and Technology, Nanjing Normal University, Nanjing 210023, People's Republic of China}
\affiliation{School of Astronomy and Space Science, 
Nanjing University, Nanjing, Jiangsu 210093, China}
\affiliation{Key Laboratory of Modern Astronomy and Astrophysics (Nanjing University), Ministry of Education, Nanjing 210093, China}

\author{Qiusheng Gu}
\affiliation{School of Astronomy and Space Science, 
Nanjing University, Nanjing, Jiangsu 210093, China}
\affiliation{Key Laboratory of Modern Astronomy and Astrophysics (Nanjing University), Ministry of Education, Nanjing 210093, China}

\author{Shiying Lu}
\affiliation{School of Physics and Astronomy, Anqing Normal University, Anqing 246133, China}
\affiliation{Institute of Astronomy and Astrophysics, Anqing Normal University, Anqing 246133, China}
\affiliation{Key Laboratory of Modern Astronomy and Astrophysics (Nanjing University), Ministry of Education, Nanjing 210093, China}

\author{Yulong Gao}
\affiliation{Research Center of Astronomy, QingHai University, Xining, 810016, PR China}
\affiliation{Department of Physics and Astronomy, QingHai University, Xining, 810016, PR China}

\correspondingauthor{Zhengyi Chen, Min Bao, Qiusheng Gu}
\email{zhengyichen2018@gmail.com}
\email{mbao@nnu.edu.cn}
\email{qsgu@nju.edu.cn}

\begin{abstract}
Wolf–Rayet (WR) stars are evolved massive stars which can drive strong stellar winds, injecting energy and momentum into the interstellar medium (ISM). However, the geometry and kinematics of WR-dominated outflows, specially in low-metallicity environments, is still poorly constrained by observations. We present a spatially resolved spectroscopic study of a WR region in a nearby dwarf galaxy, PGC\,44685, using high-resolution MEGARA IFU data from the Gran Telescopio Canarias (GTC). After decomposing the [\textsc{O iii}]~$\lambda\,5007$ emission line with narrow and broad components, we verify a WR-driven outflow with a velocity reaching up to  $20$\,km\,s$^{-1}$ relative to the systemic velocity. By use of the velocity and flux of the [\textsc{O iii}] broad component, we estimate an outflow mass of $(8.25\pm3.03)\times10^{3}\,M_\odot$ and a mass-loss rate of $(9.47\pm3.48) \times 10^{-4}\,M_\odot\,\text{yr}^{-1}$. The corresponding kinetic power and momentum injection rate are $(4.77\pm1.77)\times 10^{41}\,\mathrm{erg\,s^{-1}}$ and $(8.20\pm3.02)\times10^{28}\,\mathrm{g\,cm\,s^{-2}}$, respectively. The inferred low energy-loading efficiency ($\sim0.35\%$), together with the low metallicity of the WR region ($\sim0.1\,Z_\odot$), suggests that the system is observed in an early feedback phase in which stellar winds have not yet efficiently coupled their energy into the ISM. These results support the ability of WR feedback to shape the ISM on sub-kiloparsec scale, while these winds fail to launch galactic-scale outflows.

\end{abstract}


\section{Introduction}
Wolf-Rayet (WR) stars represent the late evolutionary stages of the most massive stars (initial mass $\gtrsim 25\,M_\odot$), typically after the hydrogen-rich outer layers being stripped by intense stellar wind or binary interaction. They are characterized by high surface temperature, broad emission lines in their spectra arising from their dense, high-velocity stellar winds, and extremely powerful stellar winds with mass-loss rates on the order of $10^{-5}\,M_\odot\,\mathrm{yr}^{-1}$ \citep{Nugis2000}, and play a crucial role in shaping the physical and chemical evolution of their host galaxies.

Through their intense feedback in the form of radiation, winds, and eventual supernova explosions, WR stars profoundly affect the surrounding interstellar medium (ISM) \citep{vink2011}. Their winds inject momentum and energy into their environment, forming bubbles and shells, compressing molecular gas, and regulating star formation on parsec to kiloparsec scale \citep{Veilleux2005}. Moreover, they are key contributors to the chemical enrichment of the ISM, particularly in helium, carbon, nitrogen, and oxygen \citep{Crowther2007}. Understanding WR wind is thus essential for constraining theoretical models of stellar evolution, supernova progenitors, and galaxy-scale feedback \citep{Sander2012, westmoquette2013ngc5253, Schaerer1999,Vacca1995}

Over the past decades, a large number of spectroscopic studies have focused on determining the global wind properties of WR stars (\citealt{Lu2024,Sander2019,del_Valle_Espinosa_2023,sirressi2024clustersuvenginesclues}), including their mass-loss rates, terminal velocities, and clumping factors. \citet{Liang2020_WR} used SDSS observations to identify WR stars in extragalactic galaxies, and further discussed their implications for the stellar initial mass function (IMF), showing that the high-mass slope of the IMF varies systematically with metallicity based on a large sample of WR regions\citep{Liang2021_WR}. These analyses have relied heavily on stellar atmosphere models that include non-local thermodynamic equilibrium (non-LTE) effects and line blanketing, such as those developed by the POWR model \citep{Hamann2004} and CMFGEN \citep{Hamann2006, Sander2012}. These models have provided robust constraints on the wind parameters of both WN and WC stars, corresponding to WR stars with nitrogen- and carbon-dominated spectral features, respectively, across different metallicity environments \citep{Vink2005_WR_winds}.

At the same time, empirical mass-loss prescriptions have been proposed as functions of stellar luminosity, metallicity, and surface composition \citep{Nugis2000}. These results are widely used in stellar evolution models, especially in predicting the fate of massive stars and the types of compact remnants \citep{vink2001}.

While the physical properties of WR stellar winds, such as mass-loss rates and terminal velocities, are relatively well constrained, their spatially resolved outflow structures still remain largely unexplored. WR stars are usually surrounded by large-scale ionized shells or so called “WR bubbles”, formed by the interaction between high-speed WR winds and the ambient ISM. A recent multiwavelength survey of WR nebulae in the Large Magellanic Cloud (LMC) revealed that about 12\% of WR stars exhibit small-scale bubbles, predominantly around WN-type stars \citep{Hung2021}. These WR bubbles display a variety of complex morphologies including shells, arcs, and clumps, indicating significant deviation from spherical symmetry and reflecting diverse evolutionary stages and environmental conditions. However, most observations of WR winds rely on spatially integrated spectroscopy, limiting detailed understanding of the internal velocity structures and anisotropies of the outflows.

In particular, the spatial distribution of outflow velocities in WR-dominated environments has rarely been investigated. As a result, our understanding of the geometry and internal kinematics of WR-driven outflows remains limited. The interaction between WR winds and the surrounding ISM, especially on small spatial scale, also remains poorly constrained. Without spatially resolved velocity mapping, it is difficult to quantify localized mass-loss rates or to evaluate hydrodynamical models of stellar feedback of the WR stars. New observational strategies are required to address these limitations.
In this work, we address these gaps by using integral field spectroscopy of PGC\,44685, a nearby star-forming galaxy, to investigate the properties of its ionized gas and stellar feedback. By analyzing high-resolution integral field unit (IFU) data obtained with MEGARA on the Gran Telescopio Canarias (GTC), we perform double-Gaussian decomposition of the [\textsc{O iii}]~$\lambda5007$ emission line across the entire field of view. This allows us to trace both the narrow component which is associated with the systemic emission, and the broad component,  tracing the outflows driven by intense WR stellar wind.
Our analysis reveals clear spatial variation in velocities of these components, indicating the presence of ionized gas outflows and asymmetric wind feedback. We construct two-dimensional maps of velocity offsets, and further estimate physical parameters such as the mass-loss rate and kinetic energy associated with the outflow. 

This work is laid out as follows. In Section~\ref{sec:data}, we describe the observational data and data reduction procedures. Section~\ref{sec:data analysis} presents the spectral fitting methodology and line decomposition. In Section~\ref{sec:discussion}, we show the spatial distribution of the outflow velocity and thus estimate the physical parameters of the outflow. Finally, the summary is given in Section~\ref{sec:conclusion}.

\section{Data and Observations}
\label{sec:data}
\subsection{GTC Data}
\label{sec:GTC}
The target of this study is PGC\,44685 (SDSS~J125958.13+020257.2; UM~533), a nearby star-forming S0 galaxy with a redshift of $z=0.00296$ \citep{Lu2024}. The galaxy hosts multiple star-forming regions, including a Wolf–Rayet (WR) region labeled C region by \cite{Lu2024}, which exhibits spectral features of prominent Blue Bump at 4686\,\AA\  and Red Bump at 5808\,\AA\ indicative of existence of WR stars.

\begin{figure}
  \centering
  \includegraphics[width=0.5\textwidth]{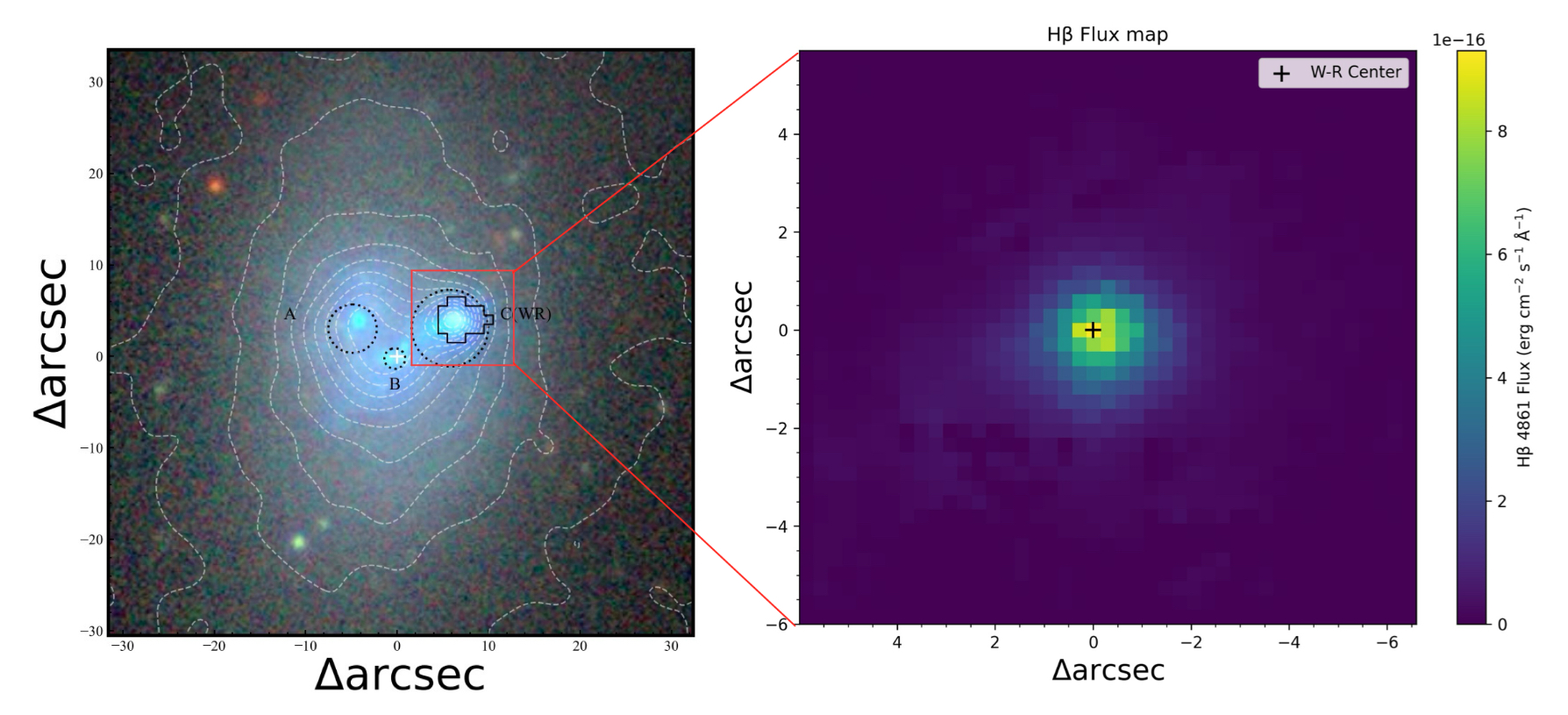}
  \caption{Optical $g$/$r$/$z$ composite image of PGC\,44685 from the DESI Legacy Imaging Surveys \citep{Dey2019}, showing the three prominent star-forming regions A, B, and C. White contours represent the corresponding $K_{\rm s}$-band emission observed with VISTA/VIRCAM. The black polygon outlines the WR region, while the red rectangle marks the field of view (FoV) of the GTC/MEGARA observations, reproduced from \citet{Lu2024}. The H$\beta$ flux map shown here is derived from the GTC IFU data and provides a detailed view of the ionized gas associated with the WR region.
  }
  \label{fig:lu_wr_map}
\end{figure}

The observation data presented in this work were obtained on February~6,~2024 (program ID \texttt{GTC36-23B}; PI: Zhengyi Chen). This observation was carried out with the MEGARA integral field spectrograph \citep{GildePaz2018} mounted on the 10.4\,m Gran Telescopio Canarias (GTC), using the Large Compact Bundle (LCB) in IFU mode. The pointing was centered on Region~C, as shown in Figure~\ref{fig:lu_wr_map}.
The total effective exposure time is $3{,}600\,\mathrm{s}$. The instrument provides a field of view of $12.5\arcsec \times 11.3\arcsec$ with a spatial sampling of $0.62\arcsec$/spaxel. The MR-B grating was employed, covering the spectral range from 4586\,\AA\ to 5024\,\AA\ at a resolution of $R \sim 12178$, which corresponds to a velocity resolution of $\sim 25$\,km\,s$^{-1}$.

The observations were carried out under an average seeing of $\sim1.2''$. The final datacube has a spatial sampling of $0.3''$ per pixel, obtained through interpolation of the fiber-based observations, and a spectral sampling of $0.11$~\AA\ per pixel.

The raw data were reduced using the official MEGARA Data Reduction Pipeline \citep{GildePaz2018}, including bias subtraction, flat-field correction, wavelength calibration, sky subtraction, and flux calibration. The final datacube is provided in physical flux units and resampled to a uniform spatial and spectral grid.

\subsection{CAHA Data}

\citet{Lu2024} presented much larger field of view optical IFU data from the 3.5\,m Calar Alto telescope with the PMAS/PPAK spectrograph. The observations used the V500 grating ($R \sim 850$, 3745-7500\,\AA) and a $1\arcsec$ spatial sampling with full dithered coverage. These data were used to identify WR star-forming region via broad He~\textsc{ii}~$\lambda4686$ emission and to estimate the gas-phase metallicity in these regions. The derived WR region map (Figure~2 of \citealt{Lu2024}), with black polygons outlining the WR region, is adopted in Section~\ref{sec:data analysis} and Section~\ref{sec:discussion}. In addition, the gas-phase metallicity ($12+\log(\mathrm{O/H})$) reported by \citealt{Lu2024} is also adopted in Section~\ref{sec:mass} to evaluate the ionized outflowing gas mass.

\section{Data Analysis and Results}
\label{sec:data analysis}
\subsection{Wolf-Rayet Region}
\label{sec:analysis}
WR stars are key tracers of recent ($\lesssim 5$ Myr) star formation and massive stellar populations \citep{Crowther2007}. In integrated galaxy spectra, WR stars are identified predominantly by the so-called ``WR blue bump'', a broad emission feature typically centered around \ion{He}{2}~$\lambda4686$, often accompanied by other lines such as [\ion{Ar}{4}]~$\lambda4711$ and \ion{Fe}{2}~$\lambda4658$ \citep{LopezSanchezEsteban2010}. 

Thanks to the integral field spectroscopy provided by MEGARA, we obtain spatially resolved spectra covering both the WR blue bump feature and the strong nebular emission lines such as H$\beta$, [\textsc{O iii}]\,$\lambda\lambda4959, 5007$. This allows us to not only identify spatial distribution of spaxels with WR features but also analyze the gas ionization. 
We apply the same procedure as described in \citet{Lu2024} to identify spaxels with WR features in MEGARA data. Specifically, the WR bump was examined within the spectral range 4650–4720\,\AA, using continuum windows (4620–4650\,\AA\ and 4750–4800\,\AA) to identify its significance following the criterion of \citet{Tresse1999}. The spaxels with WR features are contained within the WR region identified by \citealt{Lu2024}.

\begin{figure}
  \centering
  \includegraphics[width=0.5\textwidth]{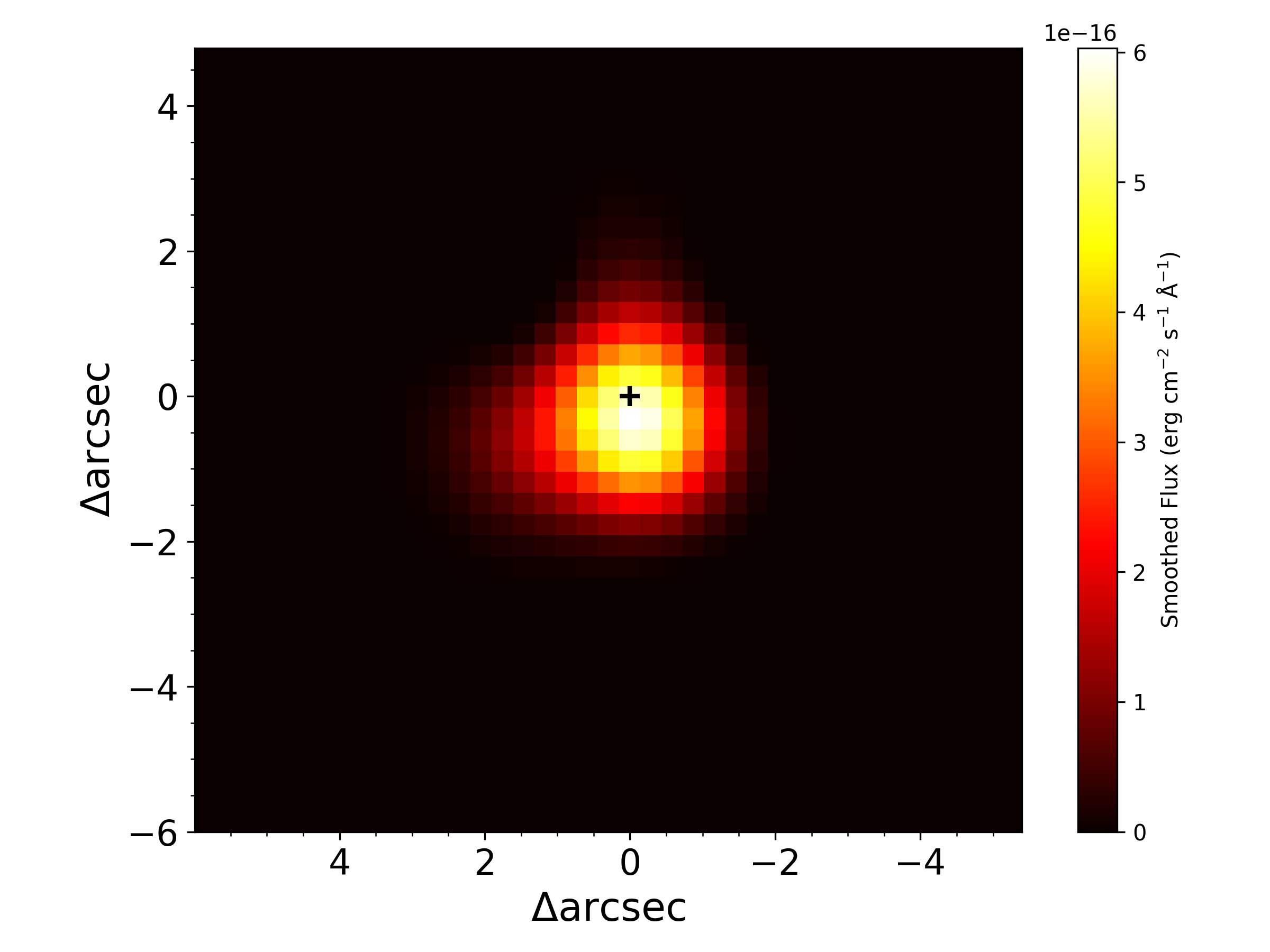}
  \caption{Flux map of the blue bump (4650–4720\,\AA). The cross represent the center of WR region. The map was convolved with a Gaussian of $\sigma = 0\farcs7$ to enhance the detection of regions showing WR features. 
}
  \label{fig:bb flux}
\end{figure}

\begin{figure*}
  \centering
  \includegraphics[width=0.98\textwidth]{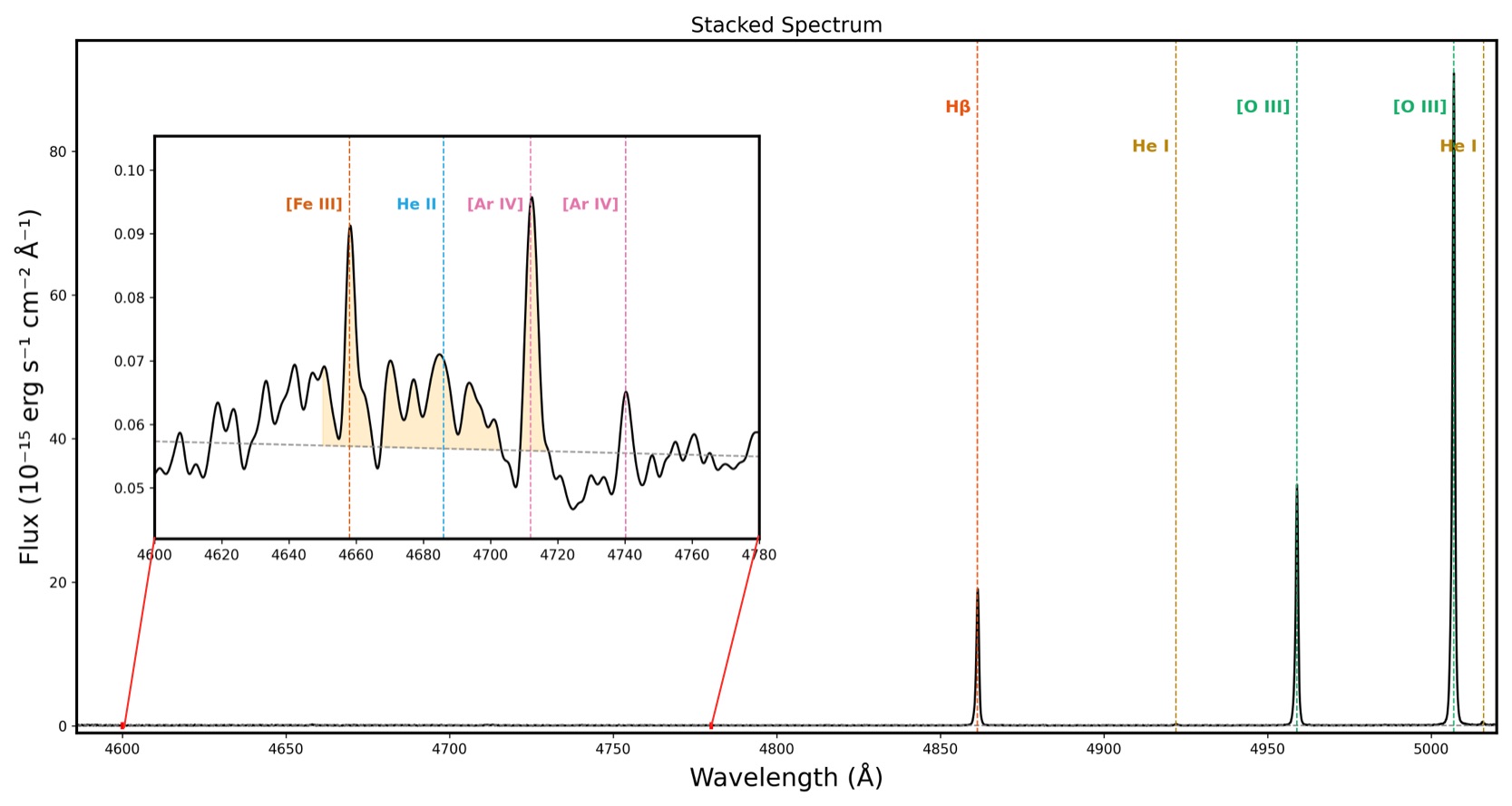}
  \caption{ The stacked WR spectrum summing all spectra with WR features. Key emission lines, including H$\beta$, [\textsc{O iii}] $\lambda\lambda4959,5007$, and He\,\textsc{i}, are marked. The upper left panel zooms in the blue bump region, where shows a broad He\,\textsc{ii} $\lambda4686$ line together with nearby Fe\,\textsc{iii} and [Ar\,\textsc{iv}] features, confirming the presence of WR stars. The dashed line represents the fitted continuum, estimated from the line-free windows, and the shaded area highlights the excess emission above the continuum within the blue bump region (4650--4720\,\AA).}
  \label{fig:spectrum}
\end{figure*}
The redshift adopted by \citealt{Lu2024} ($z = 0.00296$) is taken from the SDSS catalog, where a single systemic redshift is reported for the galaxy and commonly used to derive global quantities such as the luminosity distance. In contrast, in this work, the systemic velocity is determined directly from the MEGARA data based on the brightest [\textsc{O iii}]~$\lambda5007$ emission within the field of view. The corresponding redshift is $z = 0.00286 \pm 0.00001$, which is adopted in the subsequent analysis, including the emission-line fitting and kinematic measurements.

We construct the flux map of the blue bump by integrating the flux within the wavelength range of 4650–4720\,\AA. Following the WR feature selection procedure described in \citet{Lu2024}, WR spaxels are identified based on the detection significance of the blue bump, quantified by the parameter $\epsilon$ \citep{Tresse1999}. This quantity is defined as the ratio between the integrated flux of the blue bump and the local continuum noise, i.e., $\epsilon = F_{\mathrm{bump}} / \sigma_{\mathrm{cont}}$, where $\sigma_{\mathrm{cont}}$ is estimated from nearby line-free continuum regions. A threshold of $\epsilon > 5$ is adopted to ensure a robust detection of WR features. This criterion is used to define the WR region.

The resulting map is then convolved with a Gaussian kernel of $\sigma = 0\farcs7$ to suppress pixel-scale noise and enhance spatial coherence. As shown in Figure~\ref{fig:bb flux}, it is clear that the WR emission concentrated at the central region. The cross marks the center of the WR region, defined by the location of the maximum H$\beta$ flux. This position is consistently adopted as the WR region center throughout this analysis.
To investigate the global spectral properties of the WR region, we combine the spectra of all spaxels satisfying $\epsilon > 5$. Before stacking, each spectrum is aligned to the [\textsc{O iii}]~$\lambda5007$ velocity frame to ensure consistent co-addition of emission lines. We adopt a direct summation stacking method to maximize the signal-to-noise ratio of the intrinsically weak WR features. The stacked spectrum is used to identify WR features and to measure integrated emission-line fluxes, which are presented in Table~\ref{tab:global_fluxes}. The uncertainties of the Gaussian fitting parameters were estimated from the continuum noise by propagating the standard deviation of the residuals between the observed spectrum and the fitted continuum. We verified that these uncertainties are consistent with Monte-Carlo simulations obtained by adding Gaussian noise to the spectra and repeating the fitting procedure.

Figure~\ref{fig:spectrum} shows the resulting stacked spectrum. Prominent emission lines such as H$\beta$, [\textsc{O iii}] $\lambda\lambda4959,5007$, and several He\,\textsc{i} lines are clearly detected. In the inset at the upper-left corner, we zoom in on the blue bump region and apply mild smoothing to better highlight the features. A broad He\,\textsc{ii} $\lambda4686$ emission, together with nearby Fe\,\textsc{iii} and [Ar\,\textsc{iv}] lines, is clearly visible, confirming the presence of a WR stellar population in this region.

\begin{table}\centering\caption{Emission line fluxes measured in the WR region.}
\begin{tabular}{lcc}
\hline\hline
Line & Wavelength (\AA) & Flux (erg s$^{-1}$ cm$^{-2}$) \\
\hline
{[Fe\,III]}  & 4658.29 & $(1.60 \pm 0.16)\times10^{-16}$ \\
He\,II       & 4684.31 & $(4.67 \pm 0.37)\times10^{-16}$ \\
{[Ar\,IV]}   & 4712.18 & $(1.92 \pm 0.16)\times10^{-16}$ \\
{[Ar\,IV]}   & 4740.28 & $(6.06 \pm 1.55)\times10^{-17}$ \\
H$\beta$     & 4861.37 & $(2.14 \pm 0.01)\times10^{-14}$ \\
He\,I        & 4921.97 & $(2.51 \pm 0.16)\times10^{-16}$ \\
{[O\,III]}   & 4958.94 & $(3.57 \pm 0.01)\times10^{-14}$ \\
{[O\,III]}   & 5006.83 & $(1.02 \pm 0.01)\times10^{-13}$ \\
He\,I        & 5015.92 & $(8.73 \pm 0.16)\times10^{-16}$ \\
\hline
\end{tabular}
\label{tab:global_fluxes}
\end{table}

To explore the kinematic properties of the ionized gas in the WR region, we analyze the [\textsc{O iii}]~$\lambda5007$ emission line as a tracer of the velocity field.
In the following section, we perform Gaussian decomposition of its line profiles
to search for multiple kinematic components and signatures of outflow associated with the WR region.

\subsection{Decomposition of the [\textsc{O iii}]~$\lambda5007$  Emission Line}
\label{sec:decomposing}

\begin{figure*}[htbp]
  \centering
  \includegraphics[width=0.3\textwidth]{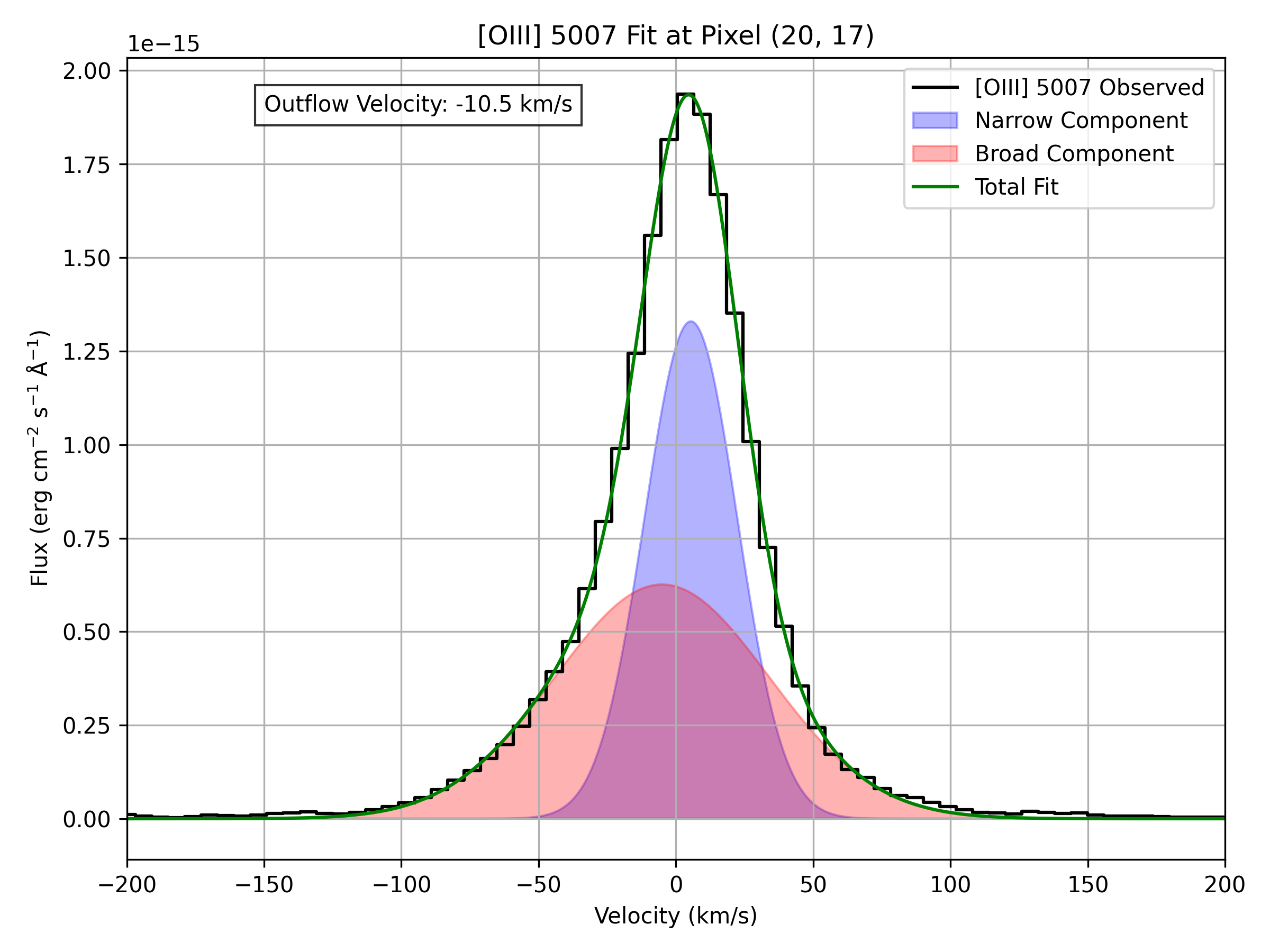}
  \includegraphics[width=0.3\textwidth]{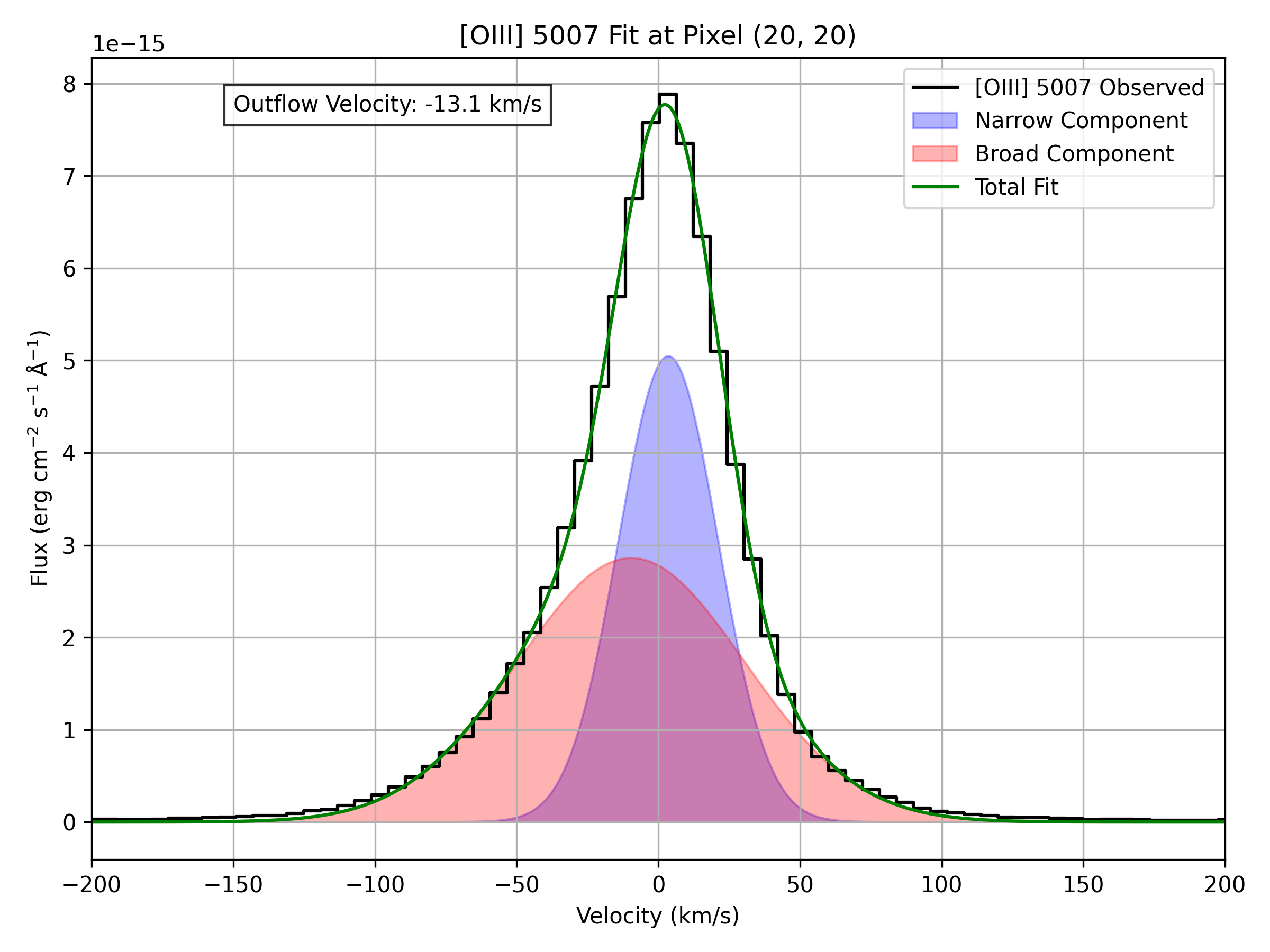}
  \includegraphics[width=0.3\textwidth]{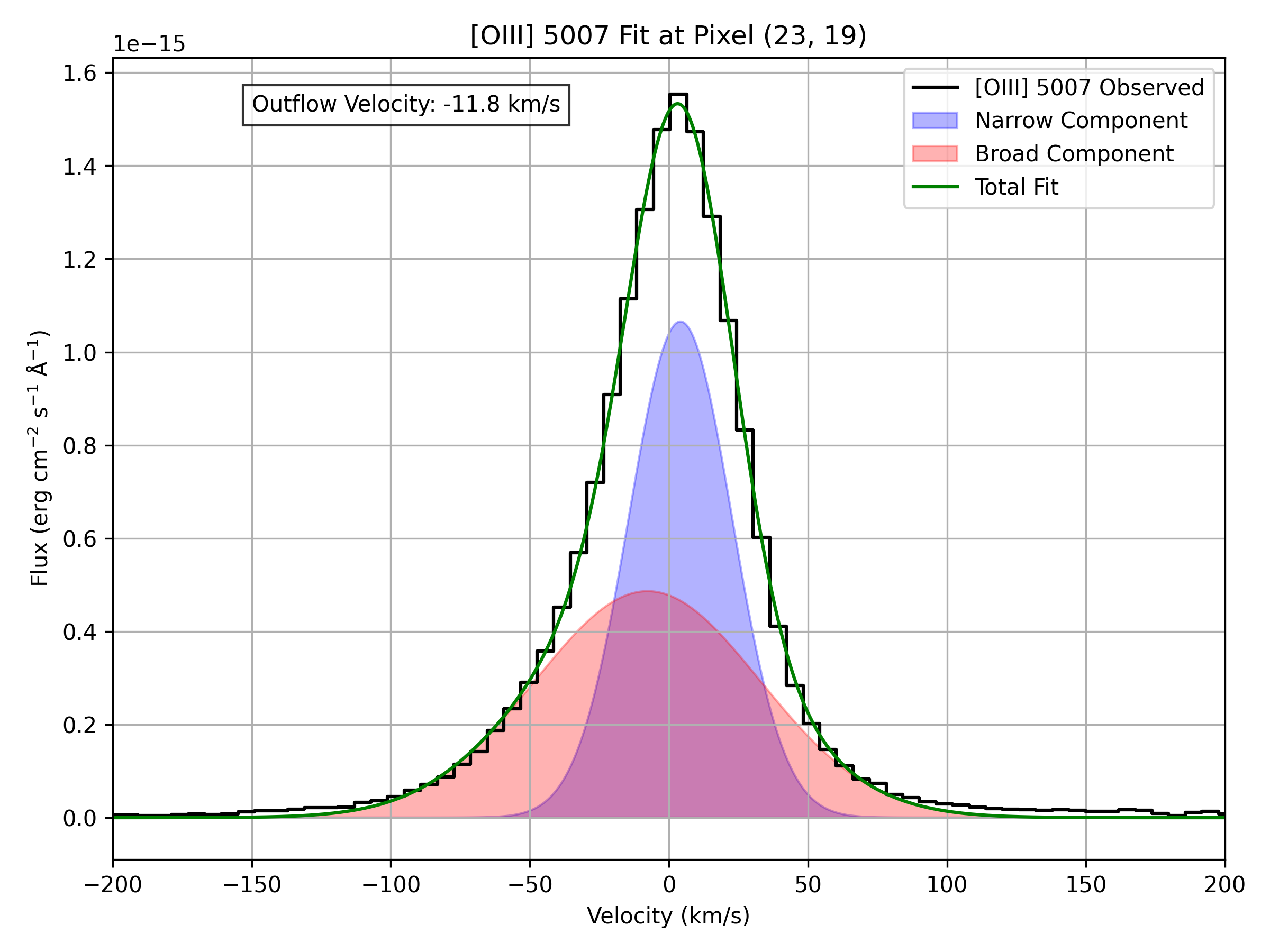}
  
  \caption{Representative examples of [\textsc{O iii}]~$\lambda5007$ line profile fits. The central panel shows the fit at the central spaxel, while the left and right panels present additional spaxels with reliable fits selected for illustration. In each panel, the black line represents the observed spectrum, the green line is the best-fit total model, and the blue and red filled curves indicate the narrow and broad Gaussian components, respectively. The derived outflow velocity, measured as the centroid offset between the broad and narrow components, is annotated in each panel.
  }

  \label{fig:nine_panel}
\end{figure*}

Across all IFU spaxels with [\textsc{O iii}]~$\lambda5007$ detected at $\mathrm{S/N} > 5$, the emission-line profiles exhibit asymmetric shapes, with extended wings that cannot be fully captured by a single Gaussian model on the blue side. 
To quantitatively assess whether an additional component is necessary, we compare the fits between single- and double-Gaussian models using the Bayesian Information Criterion (BIC; \citealt{KassRaftery1995}). The BIC is defined as $\mathrm{BIC} = \chi^2 + k \ln n$, where $k$ is the number of free parameters and $n$ is the number of data points. These criteria penalize models with extra parameters unless they lead to a significant improvement of the fit, making them appropriate to test whether a double-Gaussian model is necessary (e.g., \citealt{Liddle2007}). The double-Gaussian model yields systematically lower BIC values throughout the entire WR region, with every spaxel showing $\Delta\mathrm{BIC}>10$ and the central, highest--S/N spaxels reaching values of several thousand, consistent with the presence of multiple kinematic components, including a narrow systemic component and a broad outflow component. 
These results demonstrate that a double-Gaussian model is statistically required to describe the [\textsc{O iii}]~$\lambda5007$ emission, and we therefore adopt it in the following analysis.

To characterize the kinematics and ionized-gas properties, we perform spectral fitting of the [\textsc{O iii}]~$\lambda5007$ emission line across the field of view. All spectra were shifted to the rest frame using the redshift calculated in Section~\ref{sec:GTC}.

Flux uncertainties were estimated from the local noise measured in nearby neighboring line-free continuum windows. The continuum was modeled locally as a linear baseline, fitted within the wavelength ranges 4900--4950~\AA\ and 5010--5024~\AA. The He\,I~$\lambda4922$ and He\,I~$\lambda5016$ emission lines present in these windows were masked during the fitting to avoid contamination.

\begin{figure*}
  \centering
  \includegraphics[width=0.48\textwidth]{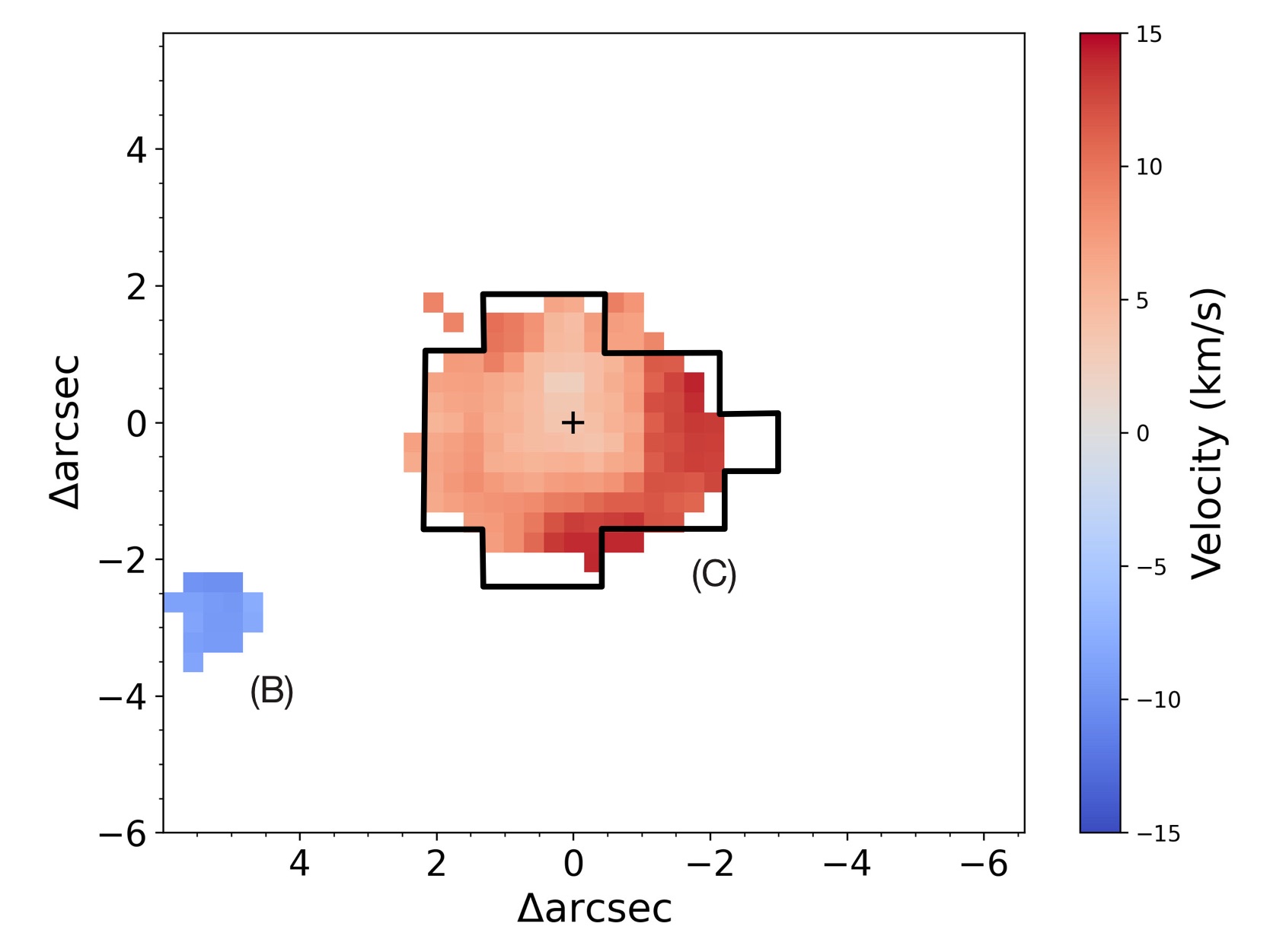}
  \includegraphics[width=0.48\textwidth]{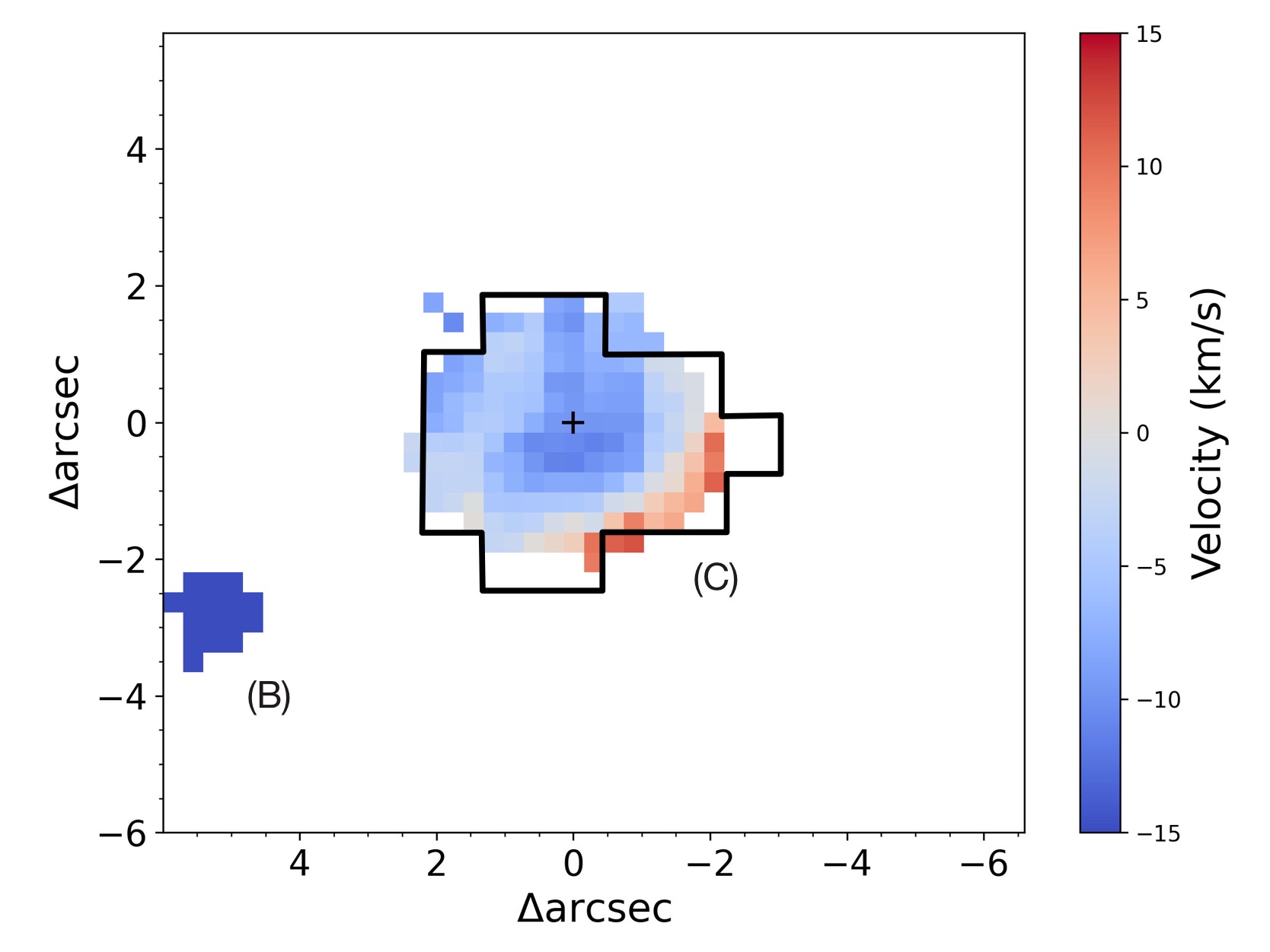}
  \caption{Decomposed velocity maps of the [\textsc{O iii}]~$\lambda5007$ emission line. 
  \textbf{Left:} Velocity map of the narrow component, which traces the systemic ionized gas. 
  \textbf{Right:} Velocity map of the broad component, which traces the outflowing ionized gas. 
  The velocity reaches its most negative value in the central region and decreases outward. 
  At the outer edges, a velocity gradient is visible, with one side redshifted and the other blueshifted, 
  suggesting the presence of a bipolar outflow possibly driven by the central WR stars.
  Velocities in both panels are measured relative to the systemic redshift of the WR region.
  The black polygons delineate the WR region identified from the blue-bump emission by \citet{Lu2024}. 
  The left-bottom part of the map corresponds to region~B, which represents another star-forming region in this galaxy. 
  The same annotations are adopted in all subsequent figures.}
  \label{fig:narrow-broad}
\end{figure*}

Each emission line was fitted with both a single-Gaussian and a double-Gaussian model:
\begin{equation}
f(\lambda) = \sum_{i=1}^{N} 
A_i \exp\!\left[-\frac{(\lambda - \lambda_i)^2}{2\sigma_i^2}\right],
\end{equation}
where $(A_i, \lambda_i, \sigma_i)$ are the amplitude, centroid, and width of each component, respectively. 
For the single-Gaussian fit, $N=1$, while for the double-Gaussian fit, $N=2$.

Emission line profiles were fitted using the Levenberg--Marquardt optimization algorithm, as implemented in the \texttt{curve\_fit} function of the \texttt{scipy.optimize} module \citep{2020SciPy-NMeth}. To ensure the robustness of the double-Gaussian model, we apply multiple physical and statistical constraints. Fits with negative fluxes, line widths $\sigma > 5$\,\AA\ or $\sigma > 300$\,km\,s$^{-1}$, or signal-to-noise ratios less than 5 (SNR~$< 5$) were discarded. Additionally, we require that the [\textsc{O iii}]~$\lambda5007$ line flux be at least twice that of the accompanying [\textsc{O iii}]~$\lambda4959$ line in the same spaxel to conform with the expected theoretical intensity ratio of $\sim2.98:1$ \citep{storey2000atomic}. 

Then we performe multiple fits for each spaxel, varying the initial guesses of the flux amplitudes and line widths for both components. Among all these trials, the solution with the lowest residual standard deviation ($\chi^2$) was selected as the final result. 

Examples of representative fits are shown in Figure~\ref{fig:nine_panel}, where the central panel corresponds to the center of the WR region. In each panel, the black curve shows the observed spectrum, the green curve denotes the best-fit total model, while the blue and red profiles represent the narrow and broad Gaussian components, respectively. All derived parameters, including fluxes, velocities, and widths were stored and used to construct two-dimensional kinematic maps in Section~\ref{sec:discussion}. 

Figure~\ref{fig:narrow-broad} shows the decomposed velocity maps of the [\textsc{O iii}]~$\lambda5007$ line.
The left and right panels represent the narrow and broad components, respectively.
The narrow component traces the systemic rotation of the ionized gas, while the broad component traces the ionized outflowing gas.
The broad component shows faster motions near the center and opposite velocity signs on either side of the central region, consistent with an ionized outflow with a bipolar geometry.
The velocity zero point in both maps is defined relative to the systemic redshift of this region, as determined in Section~\ref{sec:analysis}, and all velocities are measured with respect to this reference frame.

In both panels, the black polygons delineate the WR region identified from the blue-bump emission by \citet{Lu2024}. Both the narrow and broad velocity components obtained from the double-Gaussian decomposition are spatially concentrated within the WR region. This spatial coincidence suggests that the ionized outflow is likely associated with feedback from the central WR stellar population.

\section{Discussion}
\label{sec:discussion}

By using the decomposition of the [\textsc{O iii}]~$\lambda5007$ emission line, we can investigate the kinematic and physical properties of the ionized outflows.

To quantify the outflow kinematics, we derive the velocity offset between the narrow and broad components, which indicates the outflow velocity of the ionized gas. We can also estimate the outflow mass, mass-loss rate, momentum rate, and kinetic power of the outflows. 
These quantities provide important diagnostics of the energy budget 
and feedback efficiency of galactic winds, and are widely used in 
spatially resolved IFU studies of both star-forming galaxies 
(e.g., \citealt{Marasco2023,ReichardtChu2025}), AGN hosts (e.g., \citealt{RuschelDutra2021, nandi2025outflows}), 
and mixed samples spanning both star formation and AGN activity 
(e.g., \citealt{Arribas2014,Gallagher2019}).

\subsection{Physical Properties of Outflow}

\begin{figure}
  \centering
  \includegraphics[width=0.49\textwidth]{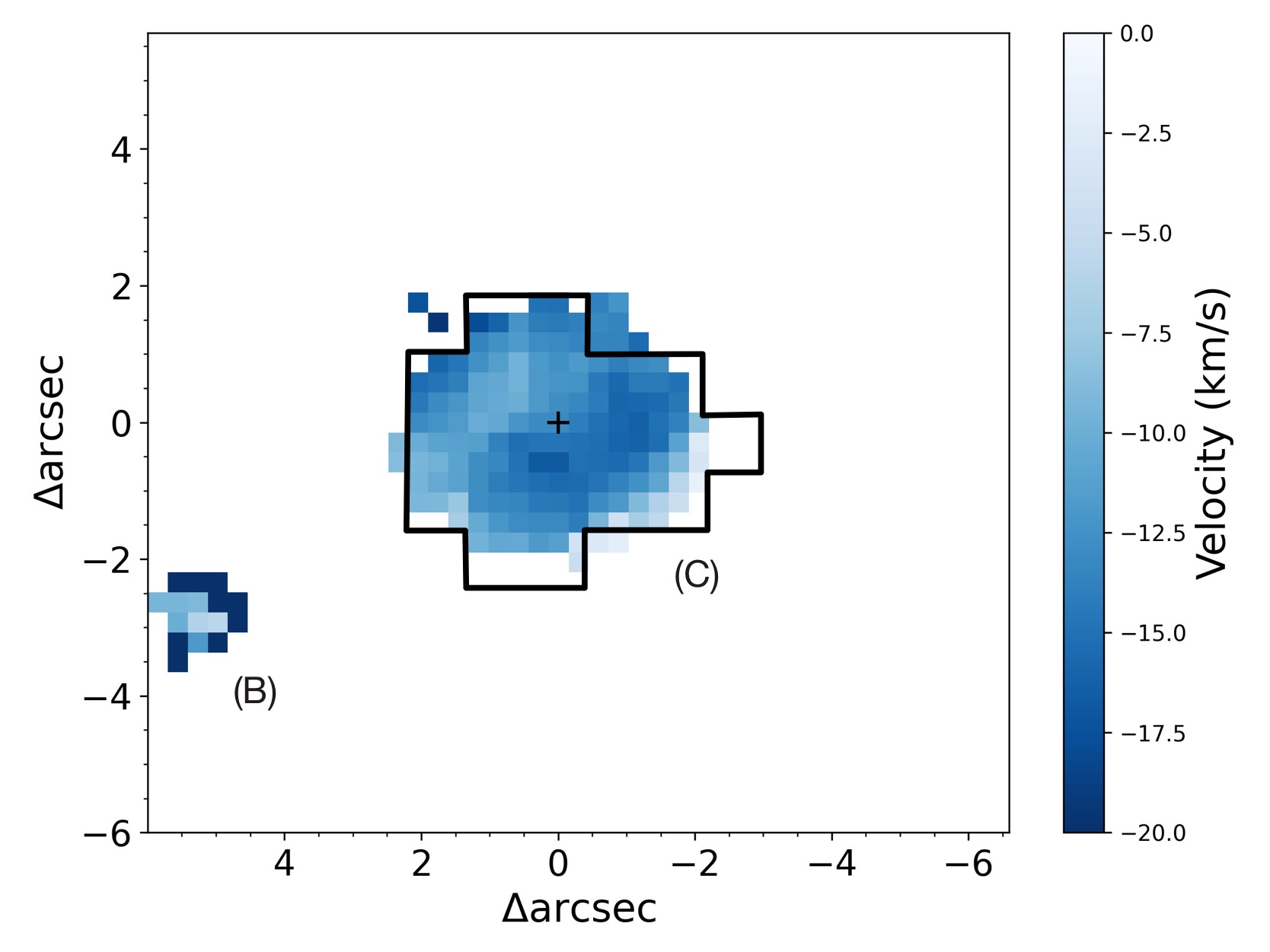}
  \caption{Outflow velocity map derived from the [\textsc{O iii}]~$\lambda5007$ emission line. 
  The outflow velocity is defined as the velocity offset between the broad and narrow components ($v_{\rm broad} - v_{\rm narrow}$). 
  Negative velocities (blueshifted) indicate outflowing gas approaching the observer.}
  \label{fig:outflow_velocity_maps}
\end{figure}

In this work, we define the outflow velocity as the velocity offset between the broad and narrow components of the [\textsc{O iii}]~$\lambda5007$ emission line:
\begin{equation}
v_{\mathrm{out}} = v_{\mathrm{broad}} - v_{\mathrm{narrow}}
\end{equation}
Negative value indicates blueshifted broad component with respect to the systemic velocity, which we interpret as gas outflow approaching us. Figure~\ref{fig:outflow_velocity_maps} shows the outflow velocity map. The black polygons delineate the WR region. Within this region, the outflow velocity field is predominantly blueshifted, supporting the presence of ionized gas outflows.

The mean outflow velocity is $-12.2 \pm 0.3$~km\,s$^{-1}$.  The velocity distribution indicates that the maximum blueshifted velocities reach approximately $-20$~km\,s$^{-1}$ within the WR region. Notably, the outflow velocities remain negative across the entire WR region, indicating a coherent and spatially persistent outflow signature. Such relatively low outflow velocities are significantly below those typically observed in star-forming galaxy samples (e.g., \citealt{Llerena2023}), where supernova (SN) feedback dominates the mechanical energy output. Recent work by \citet{SaldanaLopez2026} shows that in low-mass, low-SFR galaxies, outflows in the early stages of feedback are primarily driven by stellar winds, before the onset of significant SN activity, resulting in systematically lower velocities. Our results support a scenario in which the observed outflows are dominated by early-stage stellar feedback, prior to the development of large-scale SN-driven winds. These values are consistent with the low-velocity separation between broad and narrow components reported in WR-hosting starburst regions such as NGC\,5253 \citep{westmoquette2013ngc5253}.

It is worth noting that the nominal velocity resolution of the MEGARA instrument for our observation corresponds to $\sim 25$~km\,s$^{-1}$. However, as demonstrated by \citet{CatalanTorrecilla2023}, by employing high S/N spectra and advanced fitting techniques such as multiple repetitions, the effective precision of the measured stellar velocities can reach $\sim 9$--10~km\,s$^{-1}$. Therefore, the measured [\textsc{O iii}]~$\lambda5007$ outflow velocities in our study, with an average value of $\sim -12$~km\,s$^{-1}$, are consistent with the capabilities of MEGARA under high S/N conditions and robust fitting procedures, thereby supporting the reliability of our derived outflow velocity maps.

\begin{figure}
  \centering
  \includegraphics[width=0.49\textwidth]{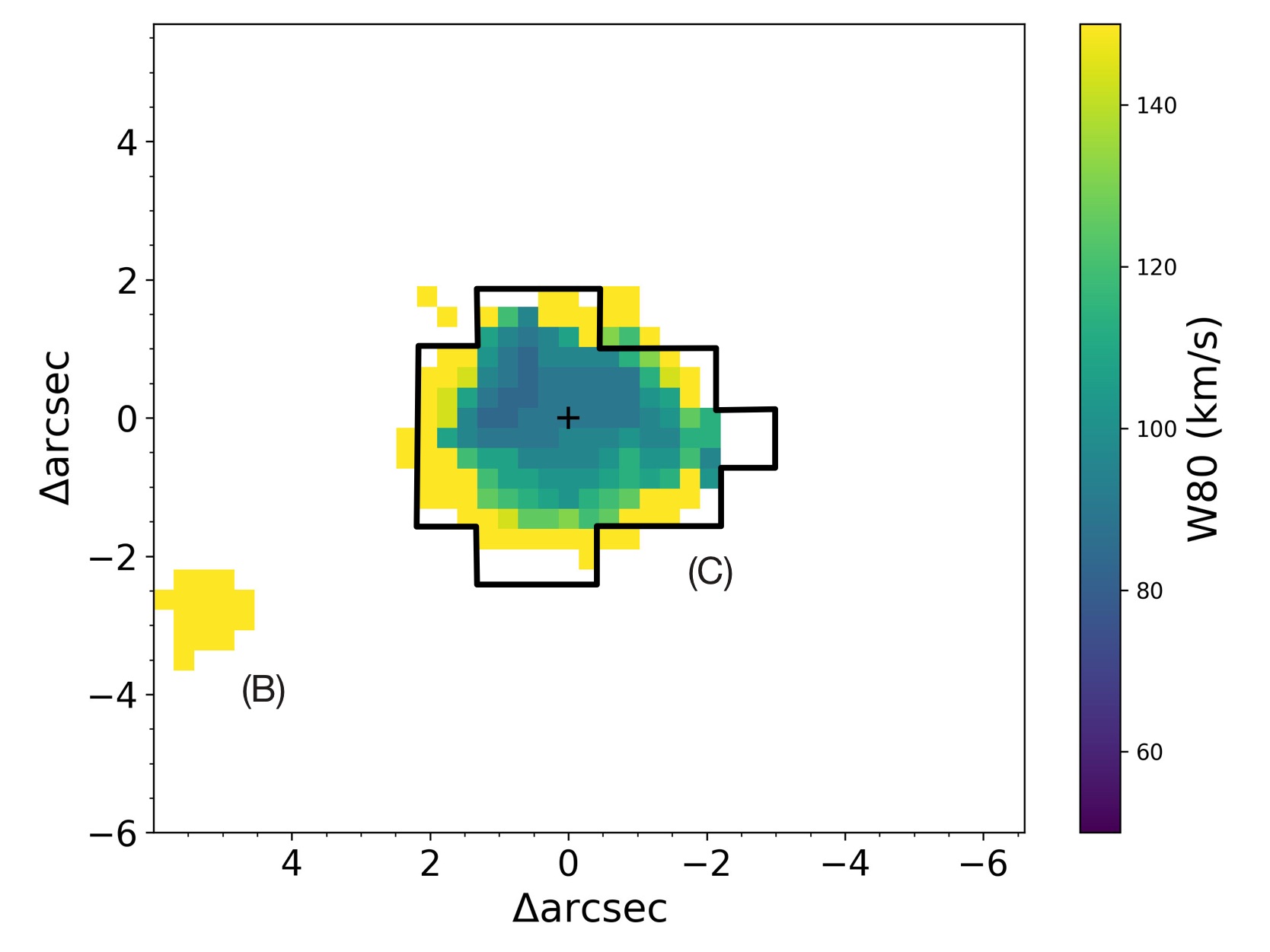}
  \caption{$W_{80}$ velocity width map derived from the [\textsc{O iii}]~$\lambda5007$ emission line. 
  An enhanced $W_{80}$ structure is observed in the outer layers of the WR region, indicating kinematically disturbed ionized gas. }
  \label{fig:w80}
\end{figure}

Figure~\ref{fig:w80} presents the $W_{80}$ map derived from the [\textsc{O iii}]~$\lambda5007$ line profile. 
The $W_{80}$ parameter, defined as $v_{90}-v_{10}$, where $v_{10}$ and $v_{90}$ correspond to the velocities enclosing 10\% and 90\% of the line flux, provides a non-parametric measure of the line broadening that is less sensitive to the detailed fitting procedure. 
An enhanced $W_{80}$ structure is evident around the outer layers of the WR region, indicating the presence of highly disturbed (or turbulent) ionized gas.

The large velocity widths in these regions are likely associated with the interaction between the WR–driven outflows and the surrounding interstellar medium. 
As the outflows driven by the WR stars expand outward, they sweep up and compress the ambient gas, forming a shell-like structure at the interface between the outflow and the surrounding ISM. 
This interaction produces enhanced turbulence and velocity dispersion in the outer layers, giving rise to the broad emission-line wings observed in the $W_{80}$ map. 
Such enhanced velocity widths are consistent with the bipolar outflow morphology seen in the velocity maps, supporting a scenario in which the ionized gas kinematics are influenced by feedback from the central WR region.

\subsection{Mass-loss Rate}
\label{sec:mass}

We then estimate the ionized gas mass associated with the outflow 
(\(M_{\mathrm{outflow}}\)) based on the [\textsc{O iii}]~\(\lambda5007\) luminosity, 
following the prescription of \citet{Carniani2015}.
The [\textsc{O iii}] emission is assumed to originate from photoionized gas 
under collisional excitation equilibrium, where the luminosity of the line 
traces the emission measure of the ionized gas 
(\(L_{\mathrm{[O\,III]}} \propto \int n_e\,n({\rm O}^{2+})\,j_{[{\rm O\,III}]}(n_e, T_e)\,dV\)). 
Combining this relation with the definition of the ionized gas mass 
(\(M \simeq m_p\,n_e\,V\)), we can obtain an analytical relation between 
the [\textsc{O iii}] luminosity and the total mass of the ionized outflowing gas:  

\begin{equation}
\label{eq:outflow_mass}
\resizebox{0.52\textwidth}{!}{$
M_{\mathrm{out}} = 0.8 \times 10^8\,  
\left( \frac{C}{10^{[{\rm O/H}] - [{\rm O/H}]_\odot}} \right)
\left( \frac{L_{\mathrm{[O\,III]}}}{10^{44}\,{\rm erg\,s^{-1}}} \right)
\left( \frac{\langle n_e \rangle}{500\,{\rm cm^{-3}}} \right)^{-1}
M_\odot$}
\end{equation}

where $C = \langle n_e^2 \rangle / \langle n_e \rangle^2$ is the clumping factor, which accounts for unresolved density inhomogeneities in the ionized gas. Given the lack of observational constraints on the clumpiness of the outflowing gas, we assume $C=1$, corresponding to a homogeneous medium, as commonly adopted in the literature \citep[e.g.,][]{Carniani2015}.

Here $L_{\mathrm{[O\,III]}}$ is the luminosity of the broad [\textsc{O iii}] component, 
$\langle n_e \rangle$ is the electron density, and $10^{[{\rm O/H}] - [{\rm O/H}]_\odot}$ represents the oxygen abundance relative to the solar value. The [\textsc{O iii}] luminosity is calculated from the observed flux of the 
[\textsc{O iii}]$\lambda5007$ emission line via

\[
L_{\mathrm{[O\,III]}} = 4\pi D_L^2 F_{5007,\,\mathrm{broad}},
\]

where \(D_L\) is the luminosity distance ($12.71~Mpc$ in our case) and 
\(F_{5007,\,\mathrm{broad}}\) is the flux of the broad component of the 
[\textsc{O iii}]~$\lambda5007$ line, obtained from the double-Gaussian fit and 
interpreted as tracing the outflowing ionized gas.

The gas-phase metallicity is adopted from \cite{Lu2024}, 
who reported a value of $12+\log(\mathrm{O/H}) = 7.71 \pm 0.16$. 
Using this value, the factor $10^{[{\rm O/H}] - [{\rm O/H}]_\odot}$ is derived accordingly, 
where the solar oxygen abundance is $12+\log(\mathrm{O/H})_\odot = 8.69$, according to \citet{Asplund2009}.

Electron density ($n_e$) was derived using the high-ionization 
[Ar\,\textsc{iv}]~$\lambda4711/\lambda4740$ diagnostic, which is 
well suited for probing the dense, high-excitation gas surrounding the WR population.  
From the integrated spectrum of the WR region, we measure a line ratio 
of 1.245. Assuming an electron temperature of $T_e = 10{,}000$~K, the 
electron density was computed using the \texttt{PyNeb} package \citep{Luridiana2015}, 
resulting in $\langle n_e \rangle \approx 1107~\mathrm{cm^{-3}}$.

Although electron densities are typically derived from the 
[S\,\textsc{ii}]~$\lambda6718/\lambda6732$ ratio, which traces low-ionization gas in \ion{H}{2} regions, it is not suitable for regions around WR stars due to the higher ionization degree. For this reason, we adopt the high-ionization 
[Ar\,\textsc{iv}]~$\lambda4711/\lambda4740$ ratio, which better reflects the physical conditions of the gas directly influenced by WR radiation and winds \citep{MonrealIbero2010}. The derived electron density of $1107~\mathrm{cm^{-3}}$ is higher than typical values obtained from other diagnostics but is consistent with results from similar WR regions. For instance, \citet{MonrealIbero2010} reported similarly high densities in such regions, attributed to the higher ionization degree associated with WR stars. Therefore, the derived value is a reasonable estimate for the ionized gas in this environment.

The mass-loss rate, \(\dot{M}_{\mathrm{outflow}}\), is then calculated assuming a bipolar outflow geometry with radius \(R_{\mathrm{outflow}}\):

\begin{equation}
\dot{M}_{\mathrm{outflow}} = \frac{M_{\mathrm{outflow}} \times v_{\mathrm{outflow}}}{R_{\mathrm{outflow}}},
\end{equation}

where \(v_{\mathrm{outflow}}\) is the characteristic outflow velocity.  The outflow radius \(R_{\mathrm{outflow}}\) is estimated from the maximum 
projected extent of the blueshifted region in our velocity maps, which 
spans approximately \(2^{\prime\prime}\), corresponding to a physical 
scale of about \(120~\mathrm{pc}\), indicating that WR feedback 
primarily affects the surrounding ISM on sub-kiloparsec scales rather than driving galaxy-scale outflows.

On average, our observations reveal prominent ionized gas outflow activity.
The total outflowing ionized gas mass is
$(8.25 \pm 3.03)\times10^{3}\,M_\odot$, and the corresponding mass outflow rate
is $(9.47 \pm 3.48)\times10^{-4}\,M_\odot\,\mathrm{yr^{-1}}$,
which is consistent with the spatial distribution of
$26 \pm 11$ WN-type stars and $3 \pm 1$ WC-type stars reported by
\citet{Lu2024}. Typically, a single WR star is expected to drive a stellar wind with a
mass-loss rate of $\sim10^{-5}\,M_\odot\,\mathrm{yr^{-1}}$
\citep[e.g.,][]{Crowther2007}.
Therefore, a population of several tens of WR stars is expected to inject
a total wind mass-loss rate of a few
$\times10^{-4}\,M_\odot\,\mathrm{yr^{-1}}$.

In addition to the above estimates, an important aspect of this system is its low gas-phase metallicity in the WR region, which is approximately $\sim0.1\,Z_\odot$. At such low metallicities, classical line-driven stellar wind theory predicts significantly reduced mass-loss rates due to the lower metal-line opacity \citep{vink2001}. This would, in principle, make it more difficult for massive stars to efficiently inject energy and momentum into the ISM, and to drive large-scale outflows.

However, WR winds differ fundamentally from those of normal OB stars. They are optically thick and can be driven by radiation pressure in deep atmospheric layers, as demonstrated by hydrodynamical atmosphere models \citep{Grafener2008}. In such optically thick conditions, radiative driving is enhanced and does not rely solely on optically thin metal lines, in contrast to classical line-driven winds. In addition, their proximity to the Eddington limit implies that continuum driving plays an important role, such that the mass-loss rate depends sensitively on the Eddington factor rather than solely on metallicity \citep{Grafener2011}. As a result, their dependence on metallicity is weaker than that of classical line-driven winds \citep{Crowther2007}. 

Therefore, even in a low-metallicity environment, WR stars may still capable of driving stellar winds and injecting material into the surrounding ISM. This scenario is consistent with the presence of the localized ionized gas outflows associated with the WR region in our observations.

\subsection{Outflow Momentum Rate and Kinetic Power}

The kinetic power and momentum rate of the outflow are computed as

\begin{equation}
\label{eq:ep}
\dot{E}_{\mathrm{kin}} = \frac{1}{2} \dot{M}_{\mathrm{outflow}} v_{\mathrm{outflow}}^{2}, 
\end{equation}

\begin{equation}
\label{eq:p}
\quad \dot{P} = \dot{M}_{\mathrm{outflow}} v_{\mathrm{outflow}}
\end{equation}
The outflow kinetic power is $(4.77\pm1.77)\times10^{41}\,\mathrm{erg\,s^{-1}}$, 
with a momentum rate of $(8.20\pm3.02)\times10^{28}\,\mathrm{g\,cm\,s^{-2}}$. 

These values are modest and consistent with expectations for WR-driven outflows operating on local scales, as both the kinetic power and momentum rate are primarily regulated by the outflow velocity. In the WR region, the measured outflow velocities are relatively low, with typical values of $\sim5$--$20\,\mathrm{km\,s^{-1}}$, which naturally leads to moderate kinetic power and momentum injection rates. It is also worth emphasizing that these quantities are evaluated within a characteristic physical scale of 
$\sim120$\,pc, corresponding to the spatial extent of the WR-dominated region.

Additionally, compared to AGN-driven outflows reported in recent studies (e.g., \citet{Xu2025}), which typically exhibit kinetic powers up to $10^{42}$~erg\,s$^{-1}$ and momentum rates exceeding $10^{32}$~g\,cm\,s$^{-2}$, the outflow energetics inferred in this work are relatively lower. A similar contrast is also found when compared to starburst galaxies, where SN-driven winds can reach kinetic powers of $\sim10^{41}$~erg\,s$^{-1}$ and momentum rates of $\sim10^{34}$~g\,cm\,s$^{-2}$ (e.g., \citet{2022ApJ...933..222X}).  This contrast reflects the different physical scales and energy sources involved. WR-driven outflows are powered by the radiatively driven winds of a limited number of massive stars, and even when integrated over a compact young cluster, the total energy and momentum injection are expected to be several orders of magnitude lower than those associated with accretion onto a supermassive black hole, or SN-dominated winds in starburst galaxies.

We further calculate the total mechanical (kinetic) energy of the outflow by:
\begin{equation}
E_{\mathrm{outflow}} = \frac{1}{2} M_{\mathrm{outflow}} v_{\mathrm{outflow}}^{2},
\end{equation}
which yields a total outflow energy of $(1.50 \pm 0.55)\times10^{49}\,\mathrm{erg}$.

To assess its physical significance, we estimate the total mechanical energy injected by the star-formation event based on the star formation rate measured in the WR region. Adopting an SFR of $0.022\,M_\odot\,\mathrm{yr^{-1}}$ \citep{Lu2024}, we compute the total output energy by scaling a \textsc{Starburst99} template corresponding to a constant star-formation rate of $1\,M_\odot\,\mathrm{yr^{-1}}$, a metallicity of $0.1\,Z_\odot$, and an age of $3$\,Myr. This approach follows the methodology presented in \citet{Hayes2023}, with further discussion in \citet{SaldanaLopez2026}.

Comparing the outflow kinetic energy to the total injected stellar energy, we obtain an energy-loading factor of $(0.35 \pm 0.13)\%$. This value is lower than those reported in studies of superluminous supernova host galaxies (SLSN hosts), where the wind energy is typically found to be about an order of magnitude below the total available energy budget (e.g., \citealt{SaldanaLopez2026}).

This difference can be understood in part due to the different tracers used to estimate the outflow mass. In this work, the outflow mass is derived from the [\textsc{O iii}]~$\lambda5007$ emission, which traces only the ionized gas phase and is therefore likely to represent a lower limit on the total outflow mass. In contrast, previous studies often estimate wind properties using absorption-line measurements that may be sensitive to a larger fraction of the outflowing material.

Additionally, the low energy-loading factor may reflect the early evolutionary stage of the system. In young star-forming regions dominated by stellar winds, a significant fraction of the injected energy is expected to remain confined within local bubbles and shells, and has not yet been efficiently transferred to the large-scale ISM. This is consistent with a scenario in which stellar feedback is still in the process of coupling to the surrounding medium \citep{Hayes2023}. Furthermore, the low metallicity of the system may contribute to the reduced feedback efficiency, since weaker stellar winds in metal-poor environments can delay the mechanical coupling between massive stars and the ISM \citep{Jecmen2023}. In this picture, stellar feedback initially produces localized bubbles and shell-like structures before the onset of SN-driven large-scale outflows.

\section{Conclusion}
\label{sec:conclusion}

 In this work, we have investigated the WR region in PGC\,44685 using MEGARA integral field spectroscopy. Based on our data processing and spectral analysis, we confirm that the WR region identified in our observations is consistent with that reported by \citet{Lu2024}. The spatial distribution of the WR features is clearly traced by the blue bump flux maps. The spatially resolved observations provide direct evidence for outflow driven by WR region. Our main results are as follows:
 
\begin{enumerate}
    \item We performed a double-Gaussian decomposition of the [\textsc{O iii}]~$\lambda5007$ emission line to separate the narrow and broad components and to derive the ionized gas outflow velocities. The broad component is found to be blueshifted relative to the systemic velocity, indicating the presence of approaching ionized gas outflows. The mean outflow velocity is $-12.2 \pm 0.3$ km\,s$^{-1}$, with the maximum blueshift reaching up to $\sim-20$ km\,s$^{-1}$. 
    
    \item Using the broad components decomposed from the [\textsc{O iii}]~$\lambda5007$ emission line, we derived the outflow properties. The outflow mass is $(8.25\pm3.03)\times10^{3}\,M_\odot$, with mass-loss rate of $ (9.47\pm3.48) \times 10^{-4}\,M_\odot\,\mathrm{yr}^{-1}$. The corresponding kinetic power is  $(4.77\pm1.77)\times10^{41}$~erg\,s$^{-1}$, and the momentum rate is $(8.20\pm3.02)\times10^{28}$~g\,cm\,s$^{-2}$. These values are consistent with the energy and momentum budget expected from the WR population within the observed region, including $26 \pm 11$ WN-type stars and $3 \pm 1$ WC-type stars. 

    \item We further compare the total kinetic energy of the outflow with the mechanical energy injected by the ongoing star formation. We find that the ratio between the outflow kinetic energy and the injected energy is only $\sim0.35\pm0.13\%$, indicating a relatively low feedback efficiency. This may reflect the early evolutionary stage of the system, in which stellar feedback is still largely confined to localized structures and has not yet efficiently coupled to the large-scale ISM. The low metallicity of the WR region ($\sim0.1\,Z_\odot$) may also contribute to delaying the mechanical coupling between stellar winds and the surrounding medium.

\end{enumerate}

\section{Acknowledgments}
We thank the anonymous referee for critical comments and instructive suggestions, which improved the content and analysis significantly. This work is supported by  the National Natural Science Foundation of China (No. 12533003, 12192222, 12192220 and 12121003). QSGU also acknowledges the science research grants from the China Manned Space Project with NO. CMS-CSST-2025-A07. M.B. acknowledges support by the National Natural Science Foundation of China, NSFC grant No. 12303009. Based on observations made with the Gran Telescopio Canarias (GTC), installed at the Spanish Observatorio del Roque de los Muchachos of the Instituto de Astrofísica de Canarias, on the island of La Palma. We acknowledge the support of the GTC staff in the observations and data reduction, in particular Antonio Cabrera Lavers and Daniel Reverte.

\bibliography{references}

\end{document}